%% file: bsource.tex
\input phyzzx
\input maggiemac
\hoffset=0.375in
\overfullrule=0pt

\def\au{{\rm AU}}

\def\au{{\rm AU}}
\def\dol{{D_{\rm ol}}}
\def\dls{{D_{\rm ls}}}
\def\sl{\ell}
\def\dv{\Delta V}
\def\dos{{D_{\rm os}}}

\def\teff{\tau_{\rm{eff}}}
\def\tmax{t_{\rm{max}}}
\def\epsil{\epsilon}

\def\dmax{\delta_{\rm{max}}}

\def\lim{{\rm lim}}

\def\kpc{{\rm kpc}}

\def\kms{{\rm km}\,{\rm s}^{-1}}

\twelvepoint
\font\bigfont=cmr17
\centerline{\bigfont Planetary Microlensing Perturbations:}
\centerline{\bigfont True Planets or Binary Sources?}
\bigskip
\centerline{\bf B.\ Scott Gaudi}
\smallskip
\centerline{Dept of Astronomy, Ohio State University, Columbus, OH 43210}
\smallskip
\centerline{e-mail gaudi@payne.mps.ohio-state.edu}
\bigskip
\centerline{\bf Abstract}
\singlespace 
A planetary microlensing event is characterized by a 
short-lived perturbation to the standard Paczy\'nski curve. 
Planetary perturbations typically last from
a few hours to a day, and have maximum amplitudes, $\dmax$, 
of $5-20\%$ of the standard curve.  There exist a subset of
binary-source events that can reproduce these main features, and
thus masquerade as planetary events.  These events require a binary
source with a small flux ratio, $\epsil \sim 10^{-2}-10^{-4}$, 
and a small impact parameter
for the fainter source, $\beta_2 \lsim \epsil / \dmax$.  The detection 
probability of events of this type is $\sim \beta_2$, and can be as
high as $\sim 30\%$; this is comparable to planetary detection rates.  
Thus a sample of planetary-like perturbations could be seriously contaminated
by binary-source events, and
there exists the possibility that completely meaningless physical parameters 
would be derived for any given perturbation.   
Here I derive analytic expressions for a binary-source event in
the extreme flux ratio limit, and use these to demonstrate the
basic degeneracy between binary source and planet perturbations.
I describe how the degeneracy can be broken by dense and accurate sampling 
of the perturbation, optical/infrared photometry, or spectroscopic
measurements.      
\bigskip
Subject Headings: gravitational lensing -- planetary systems
\smallskip
\centerline{submitted to {\it The Astrophysical Journal}: 
Jun 26, 1997}
\centerline{Preprint: OSU-TA-15/97}

\endpage
\normalspace
\chapter{Introduction}

To date more than 100 microlensing events have been detected toward the
Galactic bulge by four groups, MACHO (Alcock et al.\ 1997),
OGLE (Udalski et al.\ 1994), DUO (Alard 1996), and EROS
(Ansari et al.\ 1997).
Some of these events have been detected in real time; both MACHO and OGLE
issue `alerts', notification of ongoing
events that have been detected before the peak.  These alerts have
enabled two follow-up groups, 
PLANET (Albrow et al.\ 1996) and GMAN (Alcock et al.\ 1996), to organize
world-wide networks devoted to making densely-sampled
observations of ongoing events.  One of the main goals of these groups
is to discover planets by searching for short duration, often small, 
perturbations on the lightcurves of alerted events.  These perturbations
are the signatures of planetary events.
While standard microlensing events last from one week to a few months,
planetary perturbations are only expected to last a day or less.  Thus
the need for the intensive, nearly round-the-clock monitoring.

Previous work on planetary microlensing has focused on characterization
of the lightcurves of planetary perturbations (Wambsganss 1997),
the criteria for detection of these perturbations (Mao \& Paczy\'nski 1991;
Gould \& Loeb 1992; Bolatto \& Falco 1994; Bennett \& Rhie 1996), 
and the number of systems one might hope to detect based on these
criteria (Peale 1997). Unfortunately, mere detection of a perturbation
is not sufficient; to have any confidence that a planet has actually 
been detected, one must determine with reasonable accuracy
the physical parameters of the
planetary system that can be derived from the event,
the planet/star mass ratio, $q$ and the planet/star projected separation
in units of the Einstien ring, $y$.
Dominik (1997) discusses ambiguities in the fits of binary lenses, of
which planetary systems are a subset.  Gaudi \& Gould (1997b) demonstrated
there there exist several degeneracies which hamper the determination
of $q$ and $y$, including a severe degeneracy that can result
in an uncertainty in the derived mass ratio of a factor of $\sim 20$.

Here I discuss an additional degeneracy:
a special subset of binary {\it{source}} events
can produce lightcurves that closely resemble those produced 
by planet/star lens systems.  This subset, which I will call
extreme flux ratio binary source events, can produce standard
lightcurves with small, short duration perturbations.  These
perturbations can reproduce the gross features of 
planetary perturbations.  For a binary source event to mimic a planetary
event, the sources must have a small flux ratio, $\epsil$, and the fainter
source must pass close to the lens, with an impact parameter, $\beta_2 \lsim
\epsil/\dmax$, where $\dmax$ is the maximum fractional deviation from
the unperturbed lightcurve.  The detection probability for these events is
$\sim \beta_2$.  For $\epsil \sim 0.01$ and $\dmax \sim 0.05$, the 
probability is $\sim 20\%$.  This is comparable to the detection probability
of Jupiter-mass planets (Gould \& Loeb 1992). Thus if binary stars with 
small flux ratios are common, they could seriously contaminate a sample
of suspected planetary events.  Furthermore, for any given perturbation, 
there exists the possibility that one could derive completely meaningless
physical parameters if the perturbation
were due to a binary source rather than planet.  For these reasons,
it is essential to break this degeneracy and
determine the true cause of the perturbation (binary
source or planet).

In \S\ 2 I derive analytic expressions for the perturbation
due to a binary source in the extreme flux ratio limit. I
use these expressions in \S\ 3 to illustrate the basic degeneracy.  In \S\
4 I estimate the detection probability for extreme flux ratio binary
source events, in \S\ 5 I describe methods of 
breaking the degeneracy, and in \S\ 6 I describe how a binary
source event can be used to extract additional information about the
lens.

\chapter{Binary Source Microlensing in the Extreme Flux Ratio Limit}
\section{Basic Formalism}
The basic formalism for binary-source events has been described in
detail by Griest \& Hu (1992) for static binaries and by Han \& Gould (1997)
for rotating binaries.  Here I briefly review the general formalism,
and use this formalism to derive the equations for the 
extreme flux ratio limit.

The flux of a point source being microlensed by a point mass is
given by, $F=AF_0$, 
where $F_0$ is the unmagnified flux, and $A$ is the magnification.  
(Here I ignore any contribution from unresolved sources.) 
The magnification is a function of the distance
of the lens from the observer-source line of sight projected on the
lens plane, $u$, which is in turn a function of time:
$$ 
A[u(t)]= {{u^2 +2}\over{u(u^2+4)^{1/2}}} \rightarrow {1\over u}, \qquad 
u(t)^2 = \left[{{(t-t_0)}\over{t_e}}\right]^2 + \beta^2. \eqn\xoft
$$
The limit applies when $u \ll 1$.
Here the impact parameter, $\beta$, and $u$ are in units of the Einstein
ring,
$$
r_e^2={{4GM}\over{c^2}}{{ \dol \dls}\over{\dos}},\eqn\einrad
$$
where $M$ is the mass of the lens, and $\dol$, $\dls$, and $\dos$ are
the distances between the observer, lens and source.  The characteristic
timescale is $t_e=r_e/v$, where $v$ is the transverse velocity of the
lens relative to the observer-source line of sight.

For a binary source, the resulting lightcurve is simply a superposition
of two standard lightcurves, $F=A_1F_{0,1}+A_2F_{0,2}$ (Griest \& Hu 1992).  
Henceforth I
will assume that $F_{0,2} < F_{0,1}$ and refer to source $1$ and $2$ as
the primary and secondary, respectively.  I define
$\epsil \equiv F_{0,2}/F_{0,1}$.  The total magnification is thus
$$
A_{\rm{tot}}={{A_1+\epsil A_2}\over {1 +\epsil}}. \eqn\atot
$$
I define $b$ to be the separation of the sources projected onto the lens
plane in units of $r_e$, 
and $\theta$ to be the angle between the path of the primary
and the binary-source axis.  Assuming the binary is
static, the position of the primary
is given by equation \xoft, and the position of the secondary is,
$$
u_2^2=\left[ {{(t-t_0)}\over{t_e}}+b\cos\theta \right]^2 +(\beta_1+b
\sin\theta)^2, \eqn\xtwo
$$
where $t_0$ is the time of maximum magnification of the primary,
and $\beta_1$ is the impact parameter of the primary.
Without loss of generality, I will assume that $t_0=0$.

I now concentrate of cases such that $\epsil \ll 1$, i.e., where
the magnification of the secondary 
produces a small perturbation to the
primary lightcurve.  The fractional deviation of
such a binary-source event from the best fit single-source curve
is defined to be $\delta = (A_{\rm{tot}}-A_{\rm{bf}})/A_{\rm{bf}}$,
where $A_{\rm{bf}}$ is the best fit curve.  For $\epsil \ll 1$,
equation \atot\ implies that  $\delta \simeq \epsil A_2 / A_1$.  For $\delta$
to be significant, $A_2 \gg A_1$, and the secondary must therefore pass very
close to the lens, i.e. $|\beta_2|=|\beta_1+b\sin\theta|\ll 1$.
In this limit, equation \xoft\ implies that $A_2 \sim 1/u_2$,  
and thus when $\delta$ is significant,
$\delta \simeq \epsil/u_2A_1^{-1} $.  The maximum fractional deviation,
$\delta_{\rm{max}}\simeq \epsil/\beta_2A_1 $, 
occurs when $u_2=\beta_2$, at time  $t_{\rm{max}}=-b(\cos\theta)\,t_e$.  
The half
maximum occurs when $\delta=\dmax/2$, or $u_2=2\beta_2A_1(\beta_2)/
A_1(u_2)$.  For perturbations with short durations, the
magnification of the primary changes only very slowly during the course
of the perturbation.   Thus $A_1$ is roughly the same at $\dmax$ and
at $\dmax/2$: 
$A_1(\beta_2)\sim A_1(u_2)$.  Thus $u_2=2\beta_2$, and the full
width half maximum (FWHM) of the perturbation is 
$\teff \simeq {12}^{1/2}\beta_2 t_e$.
The equations governing binary sources in the extreme flux limit
are,
$$
\delta = {{\epsil}\over{u_2}}{1\over {A_1}}, \qquad 
\dmax ={{\epsil}\over{\beta_2}}{1 \over {A_1(t_{\rm{max}})}}, \qquad
\teff={12}^{1/2}\beta_2 t_e, \qquad
t_{\rm{max}}=-b(\cos\theta)\, t_e. \eqn\bstot
$$ 
\section{Finite Source Size Effects and Binary Rotation}

The analysis of \S\ 2.1 implicitly assumed point sources.  The point-source
approximation breaks down, however, when $u$ is ${\cal O} (\rho)$, where
$\rho$ is the radius of the source projected onto the lens plane in
units of $r_e$.  In particular, for $u \lsim \rho$, the magnification
of a finite source differs substantially from that of a point source 
(Gould 1994).  Since, for a fixed perturbation size $\dmax$,
a smaller flux ratio requires that the secondary
approach closer to the lens, there will be a lower limit on $\epsil$ below
which equation \bstot\ is no longer valid. 

Given the small flux ratios involved,
the secondary source will likely be a main-sequence star of
solar luminosity or less.  Thus I adopt a source radius of $R_{\odot}$,
which at distance of $8\, \kpc$, for a typical bulge self-lensing
event with $t_e\sim 20\, {\rm days}$, $v\sim 200 \, \kms$, and
$\dol \sim 6\, \kpc$, translates to $\rho \sim 10^{-3}$.
Thus equations \bstot\ are not valid for those events 
with $\beta_2 \lsim 10^{-3}$.  
In order to produce perturbations with $\dmax > 0.05$, 
the secondary must have an impact parameter $\beta_2 \lsim 20\epsil$. 
Thus equations \bstot\ are not valid for binary sources
with $\epsil \lsim 10^{-4}$. For flux ratios larger than this,
finite source effects can be safely disregarded, and equations \bstot\
are valid.

The effects of the rotation of the binary source for perturbations of
this type can be entirely disregarded.  The justification for this is
as follows.  To first order, the curvature of the path of the secondary 
due to rotation during the perturbation can be ignored. Thus the only 
effect is that the transverse velocity is now given by 
$v=|{\bf{v}}_{0} +{\bf{v}}_2|$, where ${\bf{v}}_{0}$ is now the transverse
velocity of the primary, and ${\bf{v}}_2$ is
the velocity of the secondary relative to primary.  The
timescale of the perturbation
will be changed, since
$\teff = 12^{1/2}\beta_2 r_e/v$. However, this effect can be reproduced by 
simply changing the value of $\beta_2$.  The observed value of $\dmax$ 
can then be reproduced by changing $\epsil$.  Therefore 
a perturbation with observables $\teff$ and $\dmax$ can
be produced by a static binary with parameters given by equation
\bstot, {\it{or}} by
a rotating binary with slightly different values of $\epsil$ and $\beta_2$.
Thus, to first order, the effect of rotation is entirely unobservable.
The second order effect is the curvature of path of the secondary during the
perturbation, which will produce effects that cannot be reproduced by
parameter variations as they can for the first order effect.  This curvature
is given by the square of the amount the binary source rotates 
during the course
of the perturbation, $\psi^2 =( 2 \pi \teff/P)^2 \simeq (22\beta_2t_e/P)^2$, 
where P is the period
of the binary source.  
Toward the galactic bulge, the typical event timescale is 
$t_e \sim 20 \, {\rm{days}}$ (Alcock et al.\ 1996).  
For bulge self-lensing events, $v\sim 200\, \kms$,
and thus $r_e \sim 2.3\, \au$.  
Using Kepler's laws, and assuming a binary-source
with separation $b=r_e$ at $8 \, \kpc$ and total mass $M=2 M_{\odot}$,
and a lens at $6 \, \kpc$, I find a binary-source separation projected
into the source plane of $3\, \au$, and a period of $P\sim 3.7\, {\rm{yr}}$.
Thus $\psi^2 \sim 0.1\beta_2^2$.
The perturbations considered here require $\beta_2 \ll 1$, and thus the
amount the binary source rotates 
during the perturbation is entirely negligible.

\chapter{Planetary Microlensing and the Basic Degeneracy}
Planetary microlensing events are a subset of binary microlensing
events with small mass ratio of the the binary, 
$q \ll 1$.  These are characterized by small perturbations
to the standard Paczy\'nski curve.  As with binary-source perturbations,
the gross features of planetary lens perturbations 
can be described by three parameters: the maximum deviation,
$\dmax$, the 
FWHM, and the time of maximum deviation, $\tmax$.
In general, $\dmax$ is a function of the geometry of the event,
the FWHM is given roughly by $\teff 
\sim q^{1/2} t_e$, where $t_e$ is the
timescale of the main lightcurve, and $\tmax$ is a
function of the planet-star projected separation in
units of Einstein ring, $y$, and the geometry of
the event, $\tmax\simeq y^{-1}(y^2-1) \cos(\phi) t_e$, where $\phi$ is
the angle between the planet-star axis and the direction of source 
motion.  Thus a planetary event is described by (Gaudi \& Gould 1997b),
$$
\teff \sim q^{1/2} t_e, \qquad \tmax\simeq y^{-1}(y^2-1) \cos(\phi) t_e,
\eqn\ppar
$$
along with $\dmax$ which specifies the exact geometry.  Here I have
ignored finite source effects.  For $q\lsim 10^{-4}$ (Neptune mass or
smaller), finite source
effects become significant; however, as I discuss in \S\ 5.1, the
severity of the degeneracy is reduced when
finite source effects are taken into consideration.  Thus, for Jupiter-mass
planetary perturbations, the following analysis is entirely applicable,
whereas for perturbations arising from planets of Neptune mass or smaller,
the analysis makes the degeneracy seem somewhat worse than it actually is.

Consider, e.g., a perturbation with observables 
$\teff = 0.03 t_e$, $\dmax =0.16$, and $\tmax=0.37 t_e$, superimposed
on a primary lightcurve with $\beta=0.37$.  Then, from equation
\ppar, a planetary
event with $q \sim 10^{-3}$, $y \sim 1.3$, and $\phi \sim 45^{\circ}$
will reproduce the observed values of $\teff$, $\dmax$, and $\tmax$.  
On the other hand, using equation \bstot, a binary source event with
$\epsil \sim 5 \times 10^{-3}$, $b \sim 0.5$, and $\theta\sim -44^{\circ}$
would also reproduce the observables.  Thus, at the level of the
gross features ($\dmax$, $\tmax$, and $\teff$), the binary
source and planetary models will provide equally satisfactory fits
to the observed perturbation.  This is the basic degeneracy, and the
example above is illustrated in Figure 1.  Note that the maximum
difference between the planetary and binary source lightcurves is
$\sim 4\%$.  

\FIG\one{
Panel (a) shows the magnification as a function of time in
units of the Einstein ring crossing time, $t_e$, for a planet/star
system (solid curve) with a mass ratio $q=10^{-3}$, a separation
in units of the Einstein ring of $y=1.3$ and angle between the
planet-star axis and direction of source motion $\phi = 45^{\circ}$,
and for a binary source system (dashed curve) with flux ratio
$\epsil=5 \times 10^{-3}$, projected separation in units of the
Einstein ring $b=0.5$ and angle between the binary source axis
and the direction of source motion $\theta=44^{\circ}$. The inset
shows a detail of the lightcurves around the time of the perturbation.
Panel (b) shows the fractional deviation from the main point-mass
point-lens lightcurve as a function of time in units of $t_e$ for
the two lightcurves in panel (a).  Both planetary (solid curve) and
binary source (dashed curve) perturbations have the same observables
$\teff = 0.03t_e$, the full width half maximum of the perturbation,
$\dmax=0.16$, the maximum fractional deviation, and $\tmax = 0.37t_e$, 
the time of maximum deviation.
}

\topinsert
\mongofigure{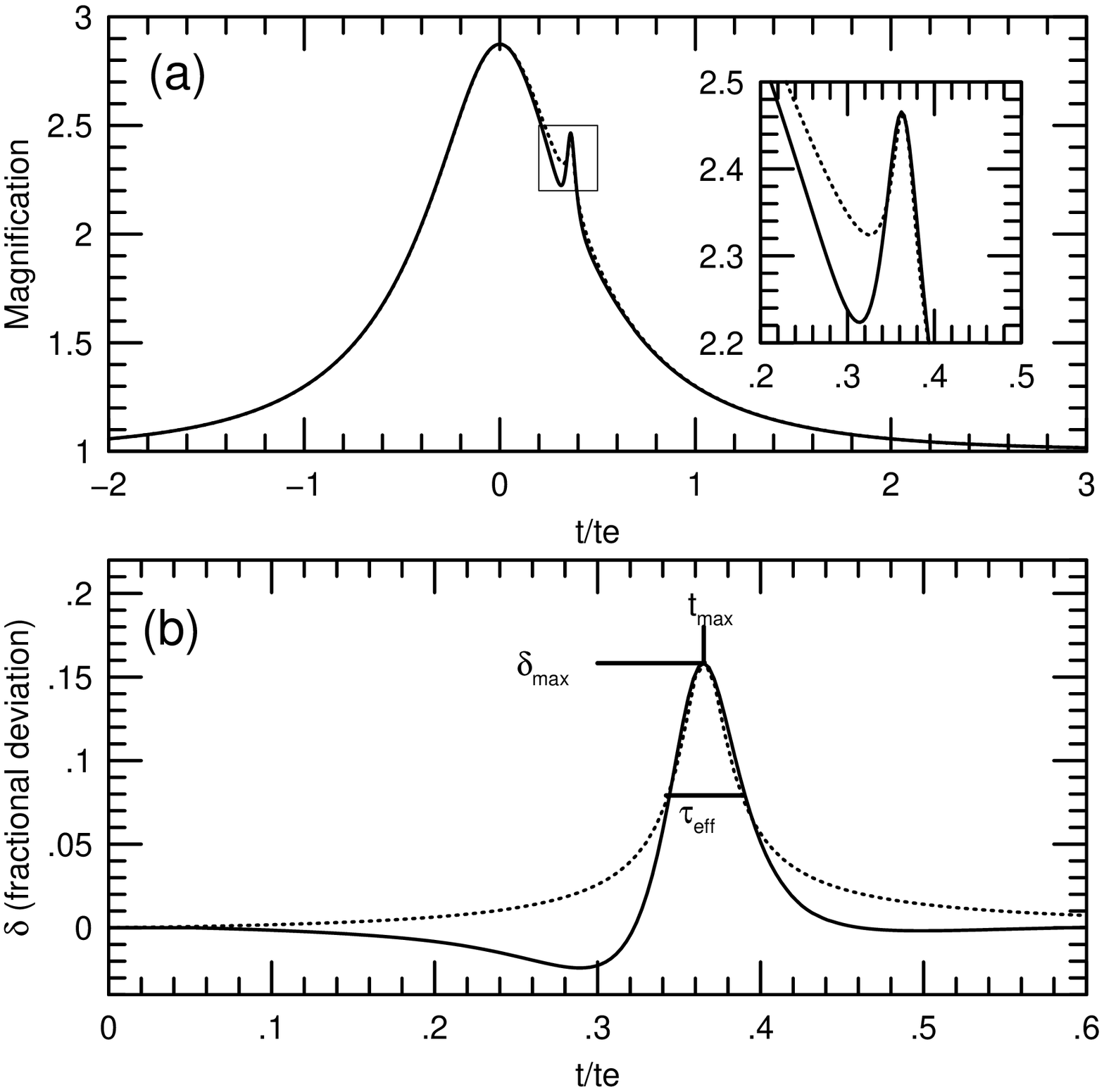}{6.4}{6.5}{7.0}
{
Panel (a) shows the magnification as a function of time in
units of the Einstein ring crossing time, $t_e$, for a planet/star
system (solid curve) with a mass ratio $q=10^{-3}$, a separation
in units of the Einstein ring of $y=1.3$ and angle between the
planet-star axis and direction of source motion $\phi = 45^{\circ}$,
and for a binary source system (dashed curve) with flux ratio
$\epsil=5 \times 10^{-3}$, projected separation in units of the
Einstein ring $b=0.5$ and angle between the binary source axis
and the direction of source motion $\theta=44^{\circ}$. The inset
shows a detail of the lightcurves around the time of the perturbation.
Panel (b) shows the fractional deviation from the main point-mass
point-lens lightcurve as a function of time in units of $t_e$ for
the two lightcurves in panel (a).  Both planetary (solid curve) and
binary source (dashed curve) perturbations have the same observables
$\teff = 0.03t_e$, the full width half maximum of the perturbation,
$\dmax=0.16$, the maximum fractional deviation, and $\tmax = 0.37t_e$, 
the time of maximum deviation.
}
\endinsert

From the example above, and the discussion in \S\ 2, it is apparent
that the basic requirements for a binary source lightcurve to mimic that of
a planetary event are a small flux ratio $\epsil$, and a specific 
geometry, i.e., one in which the fainter source passes
very close to the lens.  
More specifically, from equations \bstot, the binary 
source parameters required to reproduce an event with observables
$\teff$, $\dmax$, and $\tmax$ are,
$$
\epsil={\teff\over te}{\dmax A_1[u_1(\tmax)]\over{12^{-1/2}}}, \qquad 
b={{ \tmax \over t_e \cos\theta}},\qquad 
\theta=\tan^{-1}{\left({-\beta_1 t_e\over \tmax}\right)}, \eqn\inver
$$
where,  
as before, $A_1$ is given by equation \xoft\ evaluated at $\tmax$, and where
now $t_0=0$.  It is apparent that the value of $b$ required to fit an
observed perturbation is fixed by 
the geometry through the observables $\beta_1$ and $\tmax$.  The
required value of $\epsil$, however, depends not only on the geometry,
but also on the observed $\dmax$ and $\teff$.  Furthermore, since 
the geometry of the event affects $\epsil$ only through
$u_1(\tmax)$, and $u_1(\tmax)^2=(\tmax/t_e)^2+\beta_1^2$,
changing $\tmax/t_e$ has the same effect on $\epsil$ as changing
$\beta_1$.

\FIG\two{
Contours of the difference in magnitude between the two sources,
$\Delta V$, required to produce perturbations with the given
full width half maximum, $\teff$, and maximum fractional deviation,
$\dmax$.  The contours have spacings of $1\, {\rm mag}$. 
The solid contours are for the geometry where the primary source
has a impact parameter $\beta_1=0.3$, and the time of maximum fractional
deviation in units of the Einstein ring crossing time is 
$\tmax/t_e=0.3$.  The dotted contours are for the geometry
where {\it{either}} $\beta_1$ {\it or} $\tmax/t_e$ are smaller by $0.05$,
and the dashed contours are for the geometry where either $\beta_1$ or
$\tmax/t_e$ are larger by $0.05$.  
}
\topinsert
\mongofigure{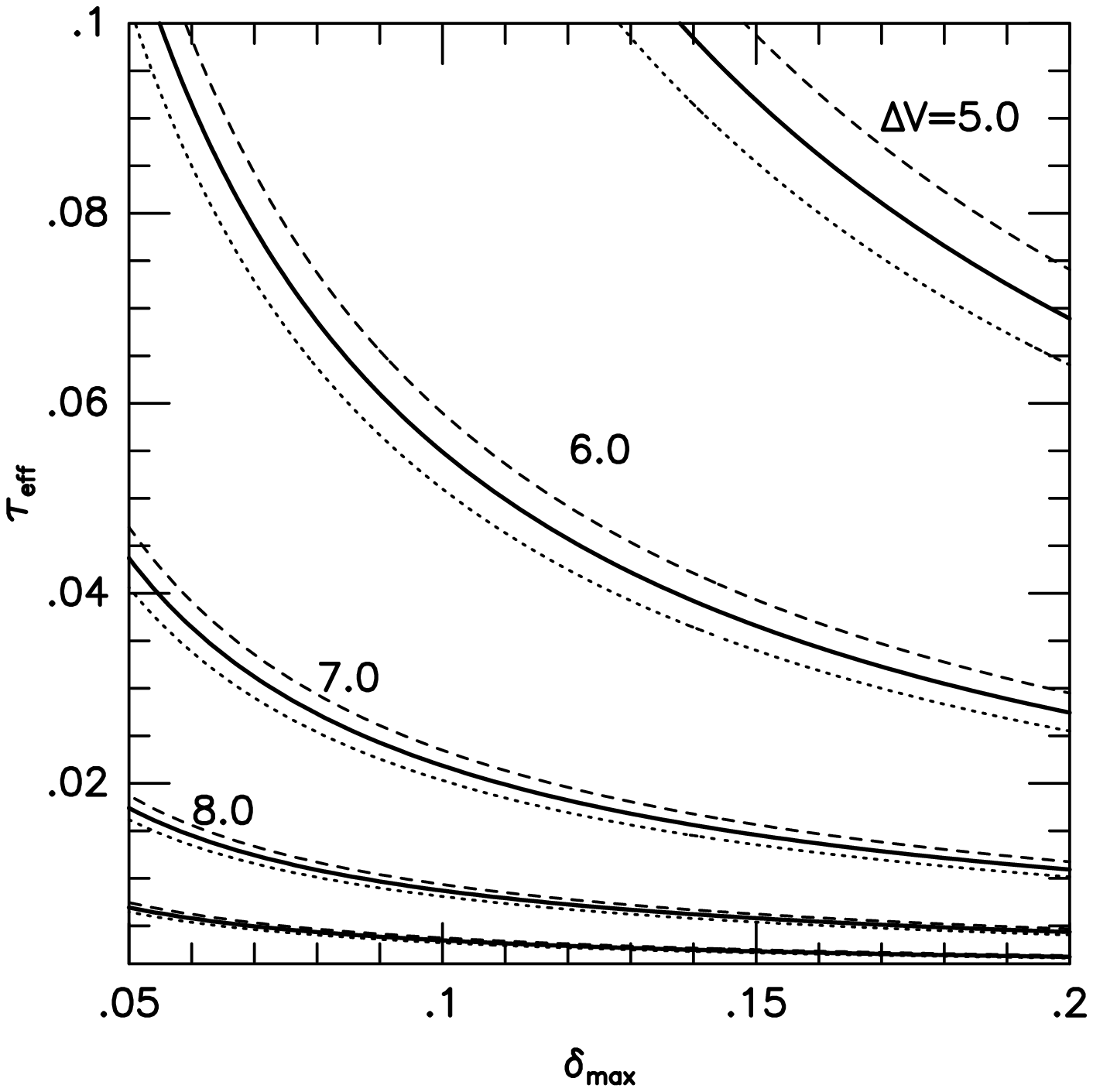}{5.5}{5.5}{5.5}
{
Contours of the difference in magnitude between the two sources,
$\Delta V$, required to produce perturbations with the given
full width half maximum, $\teff$, and maximum fractional deviation,
$\dmax$.  The contours have spacings of $1\, {\rm mag}$. 
The solid contours are for the geometry where the primary source
has a impact parameter $\beta_1=0.3$, and the time of maximum fractional
deviation in units of the Einstein ring crossing time is 
$\tmax/t_e=0.3$.  The dotted contours are for the geometry
where {\it{either}} $\beta_1$ {\it or} $\tmax/t_e$ are smaller by $0.05$,
and the dashed contours are for the geometry where either $\beta_1$ or
$\tmax/t_e$ are larger by $0.05$.  
}
\endinsert

Figure 2 shows contours of the difference in magnitude
between the two sources, $\Delta V=-2.5 \log\epsil$, required to reproduce
the given $\teff$ and $\dmax$, for three different geometries:
1) $\beta_1=0.3$, $\tmax=0.3t_e$; 2) $\beta_1$ {\it{or}} 
$\tmax/t_e$ smaller by $0.05$;
3) $\beta_1$ or $\tmax/t_e$ larger by $0.05$.  A large range of 
magnitude differences, $\dv \sim 9-5$, can produce perturbations with
$\dmax$ and $\teff$ in the ranges produced by planetary microlensing
events.  For clump giant primaries (spectral type KIII, $M_V\sim 1$ ), 
this range in $\dv$ corresponds to secondaries of spectral type 
anywhere from solar (GV) to late dwarfs (MV).

\chapter{Extreme Flux Ratio Binary Source Event Probabilities}

\FIG\three{
The fraction of binary source events that will 
be detected for the given values of the difference in
magnitude between the sources, $\Delta V$, 
as a function of the projected separation of the sources 
in units of the Einstein ring, $b$, for $\Delta V=4$ to $9$.
A binary source is considered detected when the perturbation meets the
detection criteria for
the maximum fractional deviation, $\dmax \ge 0.05$, and 
the time of maximum deviation, $\tmax/t_e \ge -1$.
}
\topinsert
\mongofigure{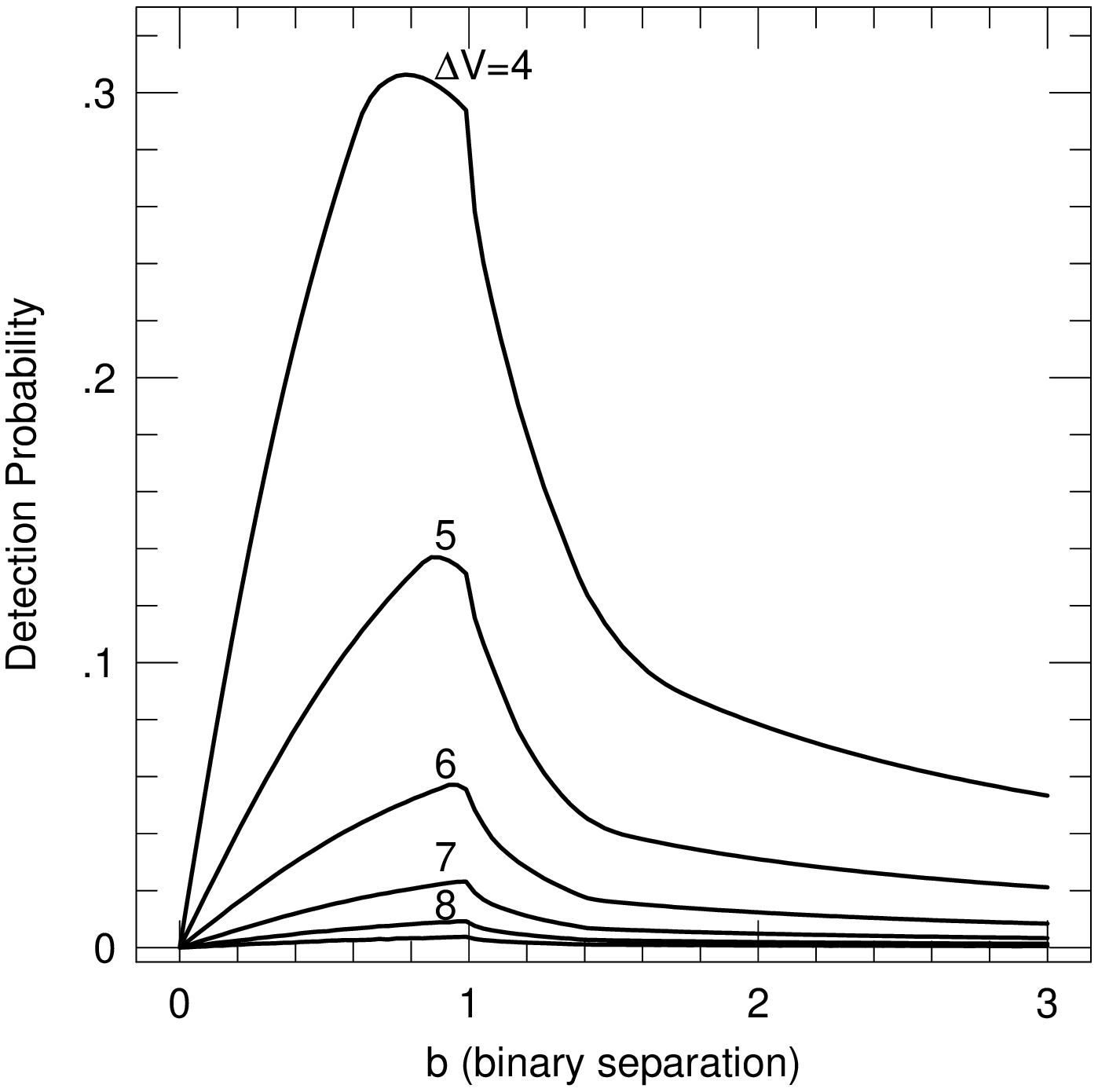}{5.5}{5.5}{5.5}
{
The fraction of binary source events that will 
be detected for the given values of the difference in
magnitude between the sources, $\Delta V$, 
as a function of the projected separation of the sources 
in units of the Einstein ring, $b$, for $\Delta V=4$ to $9$.
A binary source is considered detected when the perturbation meets the
detection criteria for
the maximum fractional deviation, $\dmax \ge 0.05$, and 
the time of maximum deviation, $\tmax/t_e \ge -1$.
}
\endinsert

For a binary source with $\epsil \ll 1$ to be detected, the lens must
pass close to the secondary. The probability that a trajectory
with any $\beta_1 \le 1$ will pass within $\beta_2$ of the secondary
is $\sim \beta_2$. Consider a binary source
with $\dv = 5$.  The secondary must have $\beta_2 \lsim 0.1$ to produce
perturbations with $\dmax \gsim 0.05$.  Thus the detection probability
for a binary source with $\dv =5$ is $\sim 20\%$.    
A more careful treatment must take into account
the fact that the magnitude of the perturbation depends on the time of the
perturbation relative to the primary lightcurve [c.f. equation \bstot].
This effect will serve to reduce the detection probability relative
to the naive estimate. 
To quantify this, I calculate, for a given $\epsil$ and $b$, the fraction
of binary source events that lead to detectable perturbations.
Although planetary events can produce a wide range of maximum deviations,
events with $\dmax < 5 \%$ are unlikely to be detected.  
I therefore  assume that the event is detected if $\dmax > 0.05$.
I place the additional constraint that $\tmax/t_e \ge -1$, 
since perturbations are unlikely to be detected
before the main event beings. 
To calculate the fraction, I integrate over $0 \le \theta < 2\pi$ and
$0 \le \beta_1 \le 1.0$.  The detection probability is simply the number of
events that satisfy the detection criteria divided by the total
number of trial events. Figure 3 shows the fraction of events
that lead to perturbations with parameters given above, for
$\dv = 4$ to $9$, and $b= 0$ to $3.0$.   For $\dv = 4$, the detection
probability can be quite high, $\sim 30 \%$.  Even for $\dv =7$,
the probability is non-negligible, and is a few percent.  

A number of authors have calculated the detection probability for
planets based on similar detection criteria.  Gould \& Loeb (1992)
found that, for Jupiter-mass planets with projected separations
$0.5 \lsim y \lsim 1.5$, the probability is $\sim 15-20 \%$. 
For Earth-mass planets with $0.5 \lsim y \lsim 1.5$, 
Bennett \& Rhie (1996) found detection probabilities
of $\sim 1-3\%$.  Since these detection probabilities are of the same order 
of magnitude as the detection probabilities for binary source perturbations 
with $\dv =4$ to $7$ and $0.5 \lsim b \lsim 1.5$, 
if binary sources with these flux ratios and projected separations are 
at least as ubiquitous as the planets the monitoring campaigns hope to detect,
they will provide a serious contaminating background.

\chapter{Breaking the Degeneracy}

As shown in \S\ 4, it is likely that binary sources will provide a 
significant contaminant in a sample of suspected planetary events.
It is therefore essential that efforts be made to resolve this degeneracy.
There are several methods to do this.

\section{Detailed Light Curves}

As is apparent from Figure 1, although a binary source and a planetary
lens can produce perturbations with the same basic features ($\teff$,
$\dmax$, and $\tmax$), the detailed light curves are dissimilar. 
In particular, during the wings of the perturbation, 
a planetary event often produces negative deviations of
a few percent, whereas binary-source perturbations produce only
positive perturbations.  For planets of $q \lsim 10^{-4}$, finite source
effects serve to increase the magnitude of the negative deviations during
the wings of the perturbation, thereby
making the binary-source and planetary perturbation more dissimilar.
Thus if one could resolve the observed lightcurve
to better than the $\sim 4\%$ level during the wings of the perturbation,
the degeneracy would be broken.  One would require dense and regular
sampling of the curve, however, since the two cases are 
significantly ($> 4\%$) different only during the first wing, and
then only for a short time ($\sim 0.1 t_e$, or $\sim {1\, \rm{day}}$ for
typical parameters).

In fact, there exist two types of planetary
perturbations: those which perturb the major image of the source
formed by the primary lens, and those which perturb the minor image.
Minor image perturbations are characterized by large ($5-20\%$)
negative deviations.  Binary source perturbations are 
therefore incompatible with minor image planetary perturbations,
and there exists no degeneracy.
 
\section{Color Information}

The most reliable way to break the degeneracy is to use color information.
If the perturbation is due to binary source, and the sources have
different colors, there will be a color change during the course
of the perturbation.  Suppose that the binary source has an (unlensed)
magnitude difference $\Delta V=(V_2-V_1)$ in $V$-band and 
$\Delta H=(H_2-H_1)$ in $H$-band.  Then I define
$\epsil_V=10^{-0.4\Delta V}$ and 
$\epsil_H=10^{-0.4\Delta H}$.  The
color change during the event is,
$\Delta(V-H)=2.5 \log{[ {A_{tot,H}/{A_{tot,V}}}]},$
where $A_{tot,V}$ and $A_{tot,H}$ are given by equation \atot,
with the appropriate $\epsil$.  Using the relation 
$\delta\simeq (A_{tot}-A_1)/A_1$, this becomes,
$$
\Delta(V-H)\simeq 2.5 \log{ {\delta_V +1}\over
{{\delta_H +1}}}. \eqn\dclra
$$
Using the relation for $\delta$ from equation
\bstot, and defining $r \equiv {\epsil_H/\epsil_V}$,
I rewrite this for the two cases $r < 1$ and
$r > 1$:
$$
\Delta(V-H)=\cases{ 
2.5 \log{ {\delta_V +1}\over{r\delta_V +1} }, & $r < 1$ \cr
2.5 \log{ {\delta_H/r +1}\over{\delta_H +1}}, & $r >1$ \cr
}.\eqn\dclrb
$$
Note that 
$2.5 \log{r}=(V-H)_2-(V-H)_1$, i.e.
the ratio $r$ is simply related to
the color difference between the secondary and 
the primary.  The maximum color change occurs at the peak of
the perturbation, and can be found by replacing $\delta_V$ in
equation \dclrb\ by $\delta_{{\rm{max}},V}$.  In particular, note that
for $r \ll 1$, 
$\Delta(V-H)\simeq 2.5 (\log_e{10})\delta_V \sim \delta_V$.
Similarly, when $r \gg 1$, $\Delta(V-H)\sim -\delta_H$. 
 Thus the largest possible color change (in magnitudes) is equal to the 
maximum ($V$ or $H$-band) fractional perturbation.

\FIG\four{
Contours of the maximum color shift $\Delta(V-H)$ in a binary
source event, as a function of the difference in colors of the two
sources, $(V-H)_2 - (V-H)_1$ and the size of the maximum fractional
deviation, $\dmax$.  The solid contours are for a shift to the blue,
$\Delta(V-H)>0$, and dotted contours are for a shift to the red,
$\Delta(V-H)<0$. If the secondary is redder than the primary,
$(V-H)_2 > (V-H)_1$, then $\Delta(V-H)<0$, and 
the maximum deviation will be in the $H$-band.  Similarly, if
the secondary is bluer than the primary, then the maximum deviation
will be in the $V$-band.
}
\topinsert
\mongofigure{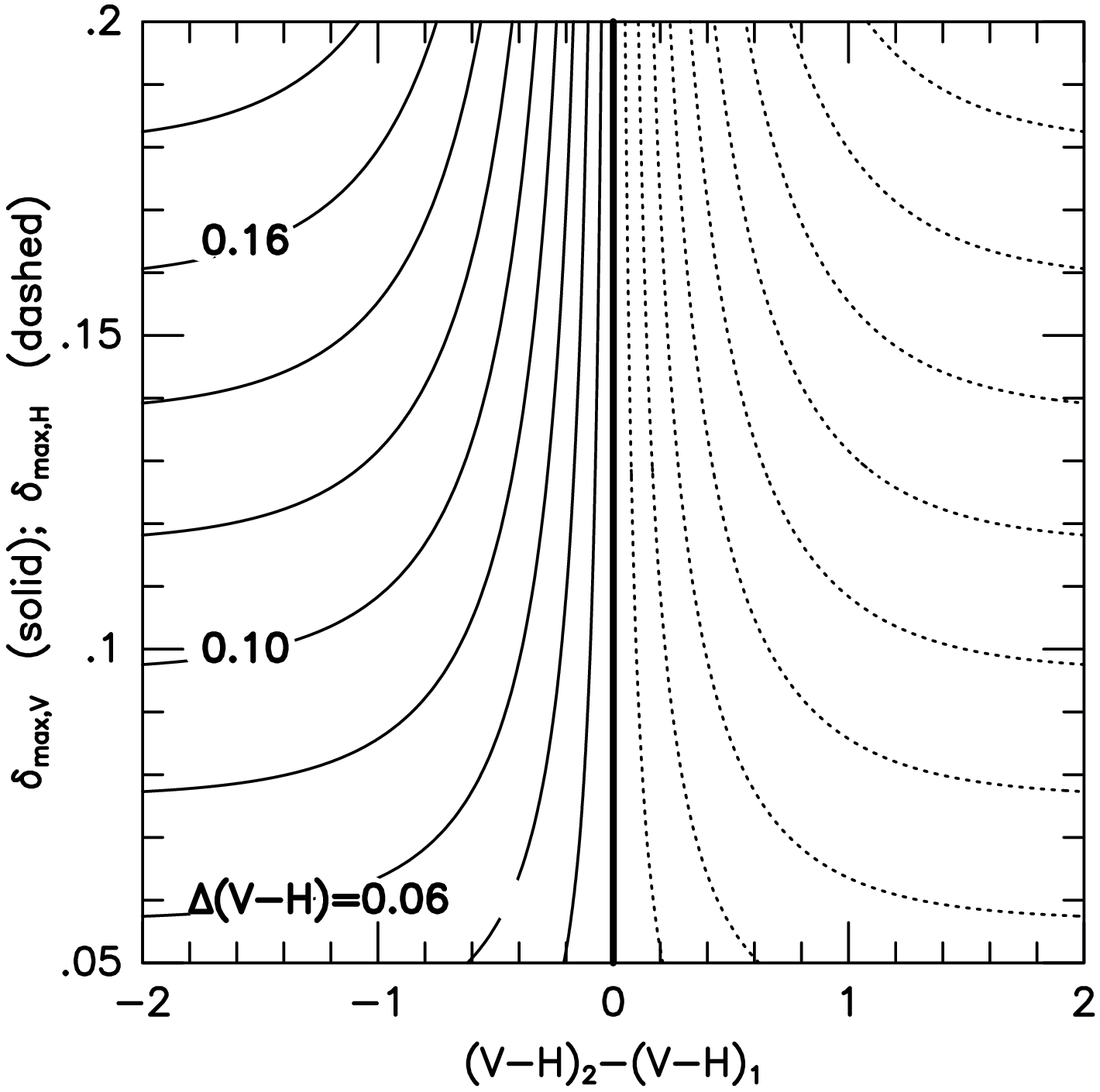}{5.5}{5.5}{5.5}
{
Contours of the maximum color shift $\Delta(V-H)$ in a binary
source event, as a function of the difference in colors of the two
sources, $(V-H)_2 - (V-H)_1$ and the size of the maximum fractional
deviation, $\dmax$.  The solid contours are for a shift to the blue,
$\Delta(V-H)>0$, and dotted contours are for a shift to the red,
$\Delta(V-H)<0$. If the secondary is redder than the primary,
$(V-H)_2 > (V-H)_1$, then $\Delta(V-H)<0$, and 
the maximum deviation will be in the $H$-band.  Similarly, if
the secondary is bluer than the primary, then the maximum deviation
will be in the $V$-band.
}
\endinsert

In Figure 4 shows contours of $\Delta(V-H)$ for $\dmax = 0.05 - 0.20$ 
and $(V-H)_2-(V-H)_1 = -2$ to $2$.  For $\sim 1\, {\rm mag}$
differences in the unlensed source colors, color changes of 
$\gsim 0.05\, {\rm{mag}}$ are produced for all measureable perturbations.
Even if the difference in source color is only $\sim 0.2\, {\rm mag}$,
substaintial ($\gsim 0.05$) color differences are produced for 
perturbations with $\dmax \gsim 0.1$.   For perspective, 
I note that for a clump giant primary ($K0III, M_V \sim 1, V-H \sim 2$), 
with a solar-type secondary ($GV, M_V \sim 5, V-H \sim 1$), 
the unlensed color difference is $\sim 1\, {\rm mag}$.  For most
binary source pairs, therefore, a significant color shift will occur
during the perturbation.

A color shift also occurs for planetary events with a small mass ratio.
The form of this shift differs significantly from that of a binary
source.  At the beginning of the planetary perturbation, 
the color first shifts to the red; during the peak, it shifts
to the blue; at the end of the perturbation, it shifts again to the
red (see, e.g., Figure 9 of Gaudi \& Gould 1997b).  This is in constrast
to binary source perturbations, where the shift is always to either
the red or blue.  Thus a color shift for a binary source can be
easily distinguished from that of a planetary event, and a measurement
of a color shift during a perturbation would allow one to unabiguously
distinguish between the two cases, and therefore break the degeneracy.

For planetary events with a large mass ratio, only a very small
color shift is produced.  Only a small color shift is produced 
for a binary source in which both sources have very similar colors.
Thus if no color shift is detected it may appear that 
the degeneracy remains.  In fact, this is not necessarily true, as
there is likely to exist a correlation between the flux ratio and
the color shift.  Assuming, for example, that the primary is known to be
a K giant.  Then, if the event is due to a binary source, the secondary
is likely to be a main sequence star.  The color-magnitude
relationship for main sequence stars translates into a relationship
between $\epsil_V$ and $r$.  This relationship, along with the value 
of $\epsil_V$ is measured from the observed lightcurve, allows one to 
estimate the expected color shift.  If the observed color shift is 
inconsisent with this estimate, then the observed perturbation cannot
be due to a binary source, and the degeneracy is broken.  

\section{Spectroscopic Methods}

If the methods suggested in \S\ 5.1 and 5.2 fail, there remain
other methods to break the degeneracy.  One possible method
is to take spectra of the source both during  and 
after the perturbation.  If the perturbation is due to a binary
source, both sources will be contributing to the
spectrum during the perturbation, whereas after the perturbation,
only the primary will contribute significantly to the spectrum.  Thus
if the binary source is a giant/dwarf pair (as it is likely to be),
then the equivilant widths of pressure sensitive spectral features
will differ between the two spectra.
Finally, one could monitor the source both 
photometrically and spectroscopically
after the event, and search for any signs of binarity. 

\chapter{Proper Motions}

If it is determined that an observed perturbation is due to a binary source
rather than planet, one can derive additional information about
the lens.  From the observed lightcurve of a binary source event, 
on can obtain the observables 
$t_e$, $\beta_1$, $\beta_2$, $t_0$, and $\tmax$.  These observables are
related to the physical projected separation, $\sl$, by (Han \& Gould 1997):
$$
\sl=\hat{r}_{e\pm} \left[ \left({{t_0-\tmax}\over{t_e}}\right)^2 + (\beta_1 \pm \beta_2)^2\right],
\eqn\refind
$$ 
where $\hat{r}_e=r_e(\dos/\dol)$ is the Einstein radius projected onto the 
source plane.  If $\sl$ can be measured by followup
spectroscopy, then $\hat{r}_e$ can be determined.  
As equation \refind\ stands, however, there exists a twofold
degeneracy in the determination of $\hat{r}_e$ due to the ambiguity
in the impact parameter difference $\Delta\beta_\pm =
|\beta_1 \pm \beta_2|$.  However, for the binary source events
considered here, $\beta_1 \gg \beta_2$, and thus 
$\Delta\beta_+\simeq \Delta\beta_- \simeq \beta_1$, and there exists no 
degeneracy.

I now discuss further the 
issue of determining $\sl$ from followup spectroscopy.
In order to determine $\sl$, the orbital elements (intrinsic physical
separation, eccentricity, true anomaly, etc.) must be determined,
and in addition the inclination angle, $i$ (c.f. Han \& Gould 1997).  
The orbital elements can be determined from a compelete radial velocity curve.
After the microlensing event, only the spectral lines of the primary
will be visible.  For a circular orbit, the maximum velocity shift of
these lines is,
$$
v_{max} = 30\, \kms \, (\sin i)\, b^{-1/2}\left({ {\cal{Q}}_M \over 
{{\cal{Q}}_M+1}}\right)^{-1/2}
\left({\hat{r}_e\over \au}\right)^{-1/2} \left( {M_1\over M_{\odot}}\right)^{1/2}.
\eqn\vmax
$$
Here ${\cal{Q}}_M = M_1/M_2$, and $M_1$ and $M_2$ are the masses of the primary
and secondary, respectively.  For a K giant primary with a solar-type 
secondary, $M_1 \sim M_\odot$ and ${\cal{Q}}_M \sim 1$.  For typical
bulge self-lensing events, ${\hat{r}_e}\sim 3\, \au$.  From Fig. 3, the
binary-source detection rate peaks at $b\sim 1$.  Thus, for typical binary
source events of this type, the expected maximum velocity shift is
$v_{max} \simeq 12\, \kms\, \sin i$.  The period of such a system is 
$P\simeq 3.7\, {\rm{yr}}$.  Excepting nearly face-on orbits,
measurement of a complete radial velocity curve
for such a system, while not trivial, is within current capabilities.
The masses of the sources are 
known approximately
from their luminosities and colors (see \S\ 5.2).
The masses can be further 
constrained if a spectrum is taken at the time of the perturbation, 
since the lines of both sources will be apparent, and
the radial velocities of these lines gives a direct measurement
of the mass ratio ${\cal{Q}}_M$.  These masses along
with the orbital elements determined from the observed
radial velocity curve determine $i$, and thus yield a complete solution
and a measurement of $\sl$.  This, combined with the event observables
$t_e$, $\beta_1$, $\beta_2$, $t_0$, and $\tmax$, yeild a measurement
of  ${\hat{r}_e}$ via equation \refind.

The fraction of events for which it is possible to measure ${\hat{r}_e}$ 
by this method is likely to be small, ${\cal{O}}(1\%)$.  I
estimate this as follows.  From Figure 3, the average 
detection rate for binary sources with $8 \lsim \dv \lsim 4$ and 
$0.5 \lsim b \lsim 1.5$ is $\sim 15\%$.   In a study of the
multiplicity of F and G stars in the solar neighborhood,
Duquennoy \& Mayor (1991) found 
that $\sim 40 \%$ of these stars had companions
with masses from $0.1$ to $1.1$ times the mass of the primary.  These
types of systems will evolve into the giant/dwarf binaries relevant
here.  Of these multiple systems, they find that $\sim 10\%$ have separations
in the range where the binary-source detection probability is high,
$0.5 \lsim b \lsim 1.5$.  Thus I estimate that 
$\sim 0.15\times 0.4 \times 0.1 \sim 1\%$ of events should display 
binary-source perturbations which can be used to measure ${\hat{r}_e}$.

The determination of $\hat{r}_e$, along with parallax information gathered from
either the Earth's motion (Gould 1992; Alcock et al.\ 1995, Buchalter
\& Kamionkowski 1997) or from a parallax satellite (Refsdal 1966;
Gould 1995; Boutreux \& Gould 1996; Gaudi \& Gould 1997a), 
yields a complete solution of the lens parameters: mass, distance,
and velocity (Gould 1996).

{\bf Acknowledgements}:
I would like to thank David Weinberg, Penny Sackett and Darren Depoy for
several stimulating discussions, 
and Andrew Gould for his careful reading of the
manuscript.  This work was supported in part by grant AST 94-20746
from the NSF.

\endpage

\Ref\alard{Alard, C.\ 1996, in Proc. IAU Symp.\ 173,
Astrophysical Applications of Gravitational Lensing, p.\ 214,
 (Eds.\ C.\ S.\ Kochanek, 
J.\ N.\ Hewitt), Kluwer Academic Publishers}
\Ref\albrow{Albrow M., et al.\ 1996, in Proc. IAU Symp.\ 173,
Astrophysical Applications of Gravitational Lensing, p.\ 227
 (Eds.\ C.\ S.\ Kochanek, 
J.\ N.\ Hewitt), Kluwer Academic Publishers}
\Ref\para{Alcock, C., et al.\ 1995, ApJL, 454, 125}
\Ref\gman{Alcock, C., et al.\ 1996, ApJL, 463, 67}
\Ref\Alcock{Alcock, C., et al.\ 1997, ApJ, 479, 119}
\Ref\Ansari{Ansari, et al.\ 1997, A\&A, in press}
\Ref\bandr{Bennett, D., \& Rhie, H.\ 1996, ApJ, 472,660}
\Ref\bandf{Bolatto, \& Falco, E.\ 1994, ApJ, 436, 112}
\Ref\bandg{Boutreux, T., \& Gould, A.\ 1996, ApJ, 462, 705}
\Ref\bandk{Buchalter, A., \& Kamionkowski, M.\ 1997, 482, 782}
\Ref\domin{Dominik, M.\ 1997, astro-ph/9703003}
\Ref\dm{Duquennoy, A., \& Mayor, M.\ 1991, A\&A, 248,485}
\Ref\gandg{Gaudi, B., \& Gould, A.\ 1997a, ApJ, 477, 152}
\Ref\gandg{Gaudi, B., \& Gould, A.\ 1997b, ApJ, 486, 000}
\Ref\exte{Gould, A.\ 1992, ApJ, 392, 442}
\Ref\goulda{Gould, A.\ 1994, ApJ, 421, L71}
\Ref\goulda{Gould, A.\ 1995, ApJ, 441, L21}

\Ref\gouldb{Gould, A.\ 1996, PASP, 108, 465}
\Ref\gandl{Gould, A., \& Loeb, A.\ 1992, ApJ, 396, 104}
\Ref\gandh{Griest, K., \& Hu, W.\ 1993, ApJ, 397, 362}
\Ref\handg{Han, C., \& Gould, A.\ 1997, ApJ, 480, 196}
\Ref\mandp{Mao, S., \& Paczy\'nski, B.\ 1991, ApJ, 374, 37}
\Ref\peal{Peale, S.\ J.\ 1997, Icarus, 127, 269}
\Ref\Refs{Refsdal, S.\ 1966, MNRAS, 134, 315}

\Ref\udal{Udalski, A., et al.\ 1994, Acta Astronomica, 44, 165}
\Ref\wamb{Wambsganss, J.\ 1997, MNRAS, 384, 172}

\refout
\endpage
\bye

%% file: maggiemac.tex
\newcount\mongocount
\mongocount=1
\def\Figure#1#2#3{
      \vbox to #3in{\hsize=#2in
        \vfil
         \includegraphics{#1}
    }
}
\def\figcap#1#2{
\vtop{\tenpoint\singlespace
\hsize=#1in\smallskip\noindent Figure\ \ \the\mongocount.\ \  #2
\global\advance\mongocount by 1\bigskip}}
\def\mongofigure#1#2#3#4#5{\centerline{\Figure{#1}{#2}{#3}
\figcap{#4}{#5}}}